\documentclass[12pt]{iopart}
\expandafter\let\csname equation*\endcsname\relax
\expandafter\let\csname endequation*\endcsname\relax
\usepackage{amssymb,graphicx,color,amsmath,bm} 
\usepackage[ bookmarks, colorlinks=true, breaklinks]{hyperref}  
\hypersetup{linkcolor=blue,citecolor=blue,filecolor=black,urlcolor=blue} 

\newcommand{\co}[1]{\left[\!\!\begin{array}{c} #1 \end{array}\!\!\right]}

\newcommand{\smatr}[1]{\begin{smallmatrix}#1\end{smallmatrix} }

\begin{document}

\title{A variational method  in the problem of screening an external charge in strongly 
correlated metals.}
\author{G. G. Guzm\'{a}n-Verri$^{1,2}$,  A. Shekhter$^{3}$, and C. M. Varma$^1$}
\address{$^1$Department of Physics and Astronomy, University of
California, Riverside, California 92521, USA \\
$^2$Materials Science Division, Argonne National Laboratory, Argonne, IL 60439 \\
$^3$Pulsed Field Facility, NHMFL, Los Alamos National Laboratory, Los Alamos, NM 87545}
\ead{gguzman-verri@anl.gov and arkady@lanl.gov}

\begin{abstract}
We describe a variational calculation for the problem of screening of a point charge in a layered correlated metal close to the Mott transition where the screening is non-linear due to the proximity to the incompressible insulating state. This analysis can robustly account for locally incompressible regions induced by external charge and gives further insights, such as overscreening in the nearest nearby metallic layers while preserving overall charge neutrality.

\end{abstract}
\date{\today}
\maketitle

Correlated metals can exhibit a large variation of electronic compressibility for small changes in the electron density. As a consequence, a local charge impurity, such as muon or dopant interstitial, may lead to a significant displacement of charge and drive the system to a local insulating (incompressible) state.  Such non-linear screening is further complicated by the layered structure of the interesting correlated metals, such as the cuprates, since an additional length scale (the layer spacing) is introduced. These aspects of the screening of a point charge in a correlated metal can be analyzed within a variational scheme.\cite{KohnSham} Though by no means exhaustive, the variational analysis allows for a phenomenological account of microscopic aspects of the physics of correlated metals which are still poorly understood. The current interest into the problem of electrostatic screening in layered correlated metals has been induced by recent muon spin relaxation experiments in underdoped metallic cuprates.\cite{Varma99a, Varma06a, Sonier01a, Sonier02a, MacDougall08a, Sonier09a, Shekhter08a, Dang10a} 

Electrostatic screening in a layered metal can be analyzed starting with a model of an equidistant stack of metallic planes at $z_n=(1/2+n)c$ separated by a dielectric medium. An external charge is placed between the two planes, at $z_n=\pm{c}/2$, at position $d$ above the center.
The electrostatic potential $\phi(\bm{r},z_n)$ is determined by screening charges $\rho(\bm{r},z_n)$ on the metallic planes via the Poisson equation  
\begin{equation}
[\phi\!\!-\!\!\phi_{\mu}]_{r,z_n}\!\!=\!\! \frac{|e|}{\epsilon} \!\!\sum_{n'}\!\!\int\!\!\!{d}^2r' \frac{\rho(\!\bm{r}'\!,\!z_{n'}\!)}{\sqrt{(\bm{r}\!\!-\!\!\bm{r}')^2\!\!+\!\!(z_n\!\!-\!\!z_{n'})^2}}, \nonumber
\end{equation}
where $\bm{r}$ is a continuous planar position. Here  $\phi_{\mu}(\bm{r},z_n)=(|e|/\epsilon)/\sqrt{\bm{r}^2+(z_n-d)^2}$ is the electrostatic potential of the external charge. Going to planar momentum variable, $\bm{\xi}$, via $\phi(\bm{\xi}) = (1/2\pi)\int{d}^2r\exp\{-i\bm{\xi}\bm{r}\}\phi(\bm{r})$ we write 
\begin{align} \label{eq:starting}
\phi(\bm{\xi},z_n)-\phi_{\mu}(\bm{\xi},z_n) = 4\pi|e| \sum_{n'} \rho(\bm{\xi},{z'}_n) \frac{e^{-\xi |z_n-{z'}_n|}}{2\xi} 
\end{align}
Due to rotational symmetry the planar momentum $\bm{\xi}$ enters only via  $\xi=|\bm{\xi}|$. Here $ \phi_{\mu}(\bm{\xi},z_n)  =  4\pi|e|{e^{-\xi|z_n-d|}}/{2\xi}$ is the potential of an external charge. Rewrite it in the form  
\begin{align} \label{eq:coulomb}
\sum_{n'}\Delta(\xi, z_n-{z}_n') ( \phi(\xi,z_n') - \phi_{\mu}(\xi,z_n') ) =4\pi e \rho(\xi,{z}_n) 
\end{align}
where  $\Delta=A^{-1}$ is  Laplacian operator on a finite stack of metallic planes defined as a matrix inverse of $A_{nn'} = e^{-\xi |z_n-{z'}_n|}/2\xi$. For a finite stack of $N$ planes it is a tri-diagonal matrix of the form,
\begin{equation}
{\Delta_{nn'}} = 
\begin{bmatrix} 
g' & f & 0 & \hdotsfor{2}  \\
f & g & f & 0 & \hdotsfor{1} \\
0 & f & g &  f & 0 \\
\hdotsfor[2]{5} \\
\hdotsfor{1}  & 0 & f & g & f \\
\hdotsfor{2}  & 0 & f & g' \\
\end{bmatrix}  
\end{equation}
where $g'=\xi(1+\coth{c\xi}), g={2\xi}\coth{c\xi}$ and $f=-\xi/\sinh{c\xi}$. Eq.~(\ref{eq:coulomb}) can be obtained as a condition for stationarity of the functional 
\begin{eqnarray}
S\!\!=\!\!\int\!\!\!\frac{d^2\xi}{(2\pi)^2}\frac12\sum_{nn'}\psi(\xi,z_n)\Delta(\xi, z_n-z_n')\psi(\xi,z_n') & \nonumber \\
&\hspace{-1cm}+S_F(\{\phi(r,z_n)\})
\label{eq:functional1}
\end{eqnarray}
with $\phi$ as an independent variable. Here $\psi(\xi,z_n)=\phi(\xi,z_n)-\phi_{\mu}(\xi,z_n)$. The screening charge density is defined via the second term by  $4\pi|e|\rho(r,z_n)= - dS_F/d\phi(r,z_n)$. For a metallic plane characterized by linear compressibility we have $\rho=-(\kappa/2\pi|e|)\phi$ where $\kappa= (2\pi\, |e|^2/\epsilon) (dn/d\mu)$ is the inverse screening length of the two-dimensional metal\cite{Ashcroft} and 
\begin{align}
S_F^0 =\int\frac{d^2\xi}{(2\pi)^2} \frac12 \sum_n 2\kappa \phi(\xi,z_n)^2
\end{align}
The form of $S_F$ for non-linear screening will be discussed later. We will take the simple view that for small displaced charge the metallic plane has a large and fixed charge compressibility and therefore the screening is linear. Thus in the situation where most of the screening of external charge is done by the two nearest planes, $z_n=\pm{c/2}$, the screening charge in the rest of metallic planes is small and therefore can be accounted within linear screening regime. In this situation one has to consider the effects of non-linear screening in the two nearest metallic planes only. 

To obtain a functional that depends on the potentials in the two nearest planes only\cite{Feynman} we rewrite Eq.~(\ref{eq:functional1}) in the matrix form (the $\xi$-integrals are implied) 
\begin{align}\label{eq:blocks}
S =&\frac12\left[\begin{array}{ccc}
\Psi_{+} &
\Psi_{0} &
\Psi_{-} 
\end{array}
\right]
\left[\begin{array}{ccc}
 \Delta_{++}  & \Delta_{+0}& 0 \\
\Delta_{0+} & \Delta_{00} & \Delta_{0-} \\
0 & \Delta_{-0} & \Delta_{--} 
\end{array}
\right]
\left[\begin{array}{c}
\Psi_{+} \\
\Psi_{0} \\
\Psi_{-} 
\end{array}
\right] \notag\\
&~~~~~~~~~~~~~~~~~~~~~~~~~~~~~~~~~~~~~~~~ 
+\frac12 [ (2\kappa) (\Psi_+ + \Phi^{\mu}_+)^2 + (2\kappa) (\Psi_- + \Phi^{\mu}_-)^2 ]
\end{align}
We have kept the (linear) compressibility term for all planes except the two in the center, $z_n=\pm{c/2}$. Here $\Psi=\{\psi(\xi,z_n)\}_n$ is a vector planar index as an index. Similarly,  $\Phi_{\mu}=\{\phi_{\mu}(\xi,z_n)\}_n$. We separate out the two nearest planes $\Psi_0=\{\psi(\xi,z_{+1/2}), \psi(\xi,z_{-1/2})\}$, the rest of planes above, $\Psi_+$, and the rest of planes below, $\Psi_-$ (and similarly for $\Phi_{\mu}$).  The 2x2 block, $\Delta_{00}$, multiplies the two planes near the external charge. The block $\Delta_{+0}^{(2)}$  is $2 \times \frac{N-2}2$  block with only one non-zero element in correspondence with the tri-diagonal structure of matrix $\Delta$. Omitting $\Psi_0$-independent terms, we write 
\begin{eqnarray}\label{eq:matrixdelta1}
S=&\frac12\Psi_0^{\dagger}[\Delta_{00}+\Delta_1]\Psi_0 -\frac12 [\Psi_0^{\dagger} \tilde{\Phi}_{\mu} + \tilde{\Phi}_{\mu}^{\dagger} \Psi_0],\\
\Delta_1 =& -\Delta_{0+} [2\kappa + \Delta_{++}]^{-1} \Delta_{+0} + (+\leftrightarrow-), \nonumber \\
\tilde{\Phi}^{\mu} =& 2\kappa  \Delta_{0+} [2\kappa + \Delta_{++}]^{-1} \Phi^{\mu}_{+} 
+ (+\leftrightarrow-) \nonumber
\end{eqnarray}
which defines the bilinear and linear terms in the functional for two central planes. The first term in the expression for $\Delta_1$ has a form $\left[\smatr{B & 0 \\ 0 & 0 }\right]$ with $B=(-\xi/\sinh{c\xi})^2[2\kappa+\Delta_{(++)}]^{-1}|_{N,N}$, (here $N$ is the dimension of the $\Delta_{++}$ matrix). We define matrix $D$ via  $2\kappa+\Delta_{(++)} = [2\xi\coth{c\xi} +2\kappa] D $. It is equal to $1$ on the main diagonal and $-1/(2\alpha)$ on the two near diagonals,  $\alpha(\xi)=\cosh{c\xi}+(\kappa/\xi)\sinh{c\xi}$. Taking the limit $N\rightarrow\infty$, ignoring $\alpha'$, and using the identity $[D^{-1}]_{NN}=(1/2\pi)\int_{-\pi}^{\pi}d\theta{2\alpha\sin^2\theta}/[\alpha - \cos\theta]$ we obtain $B = \left(\xi/\sinh{c\xi}\right) [\alpha-\sqrt{\alpha^2-1}]$. Adding the $(--)$ block in Eq.~(\ref{eq:matrixdelta1}), 
\begin{align}
{\Delta_{00}}+{\Delta_{1}} = 
\begin{bmatrix} 
2{\xi}{\coth{c\xi}} -B & -\frac{\xi}{\sinh{c\xi}} \\
-\frac{\xi}{\sinh{c\xi}} & 2{\xi}{\coth{c\xi}} -B \\
\end{bmatrix}
\end{align}
The linear terms in Eq.~(\ref{eq:matrixdelta1})  are evaluated similarly,  
\begin{eqnarray}\label{eq:sum21}
\tilde{\Phi}_{\mu} =  \begin{bmatrix} 
-C_{+c/2} \\
-C_{-c/2} 
\end{bmatrix} \,,\\
C_{\pm c/2} =  \frac{2\kappa}{2\alpha} 
\times \sum_{k=1}^{\infty} [D^{-1}]_{k,1}\phi_{\mu}(z_n=\pm{c}(k+1/2))
\hfill \nonumber
\end{eqnarray}
We obtain (here $\phi_{\pm} = \phi(z_n=c/2)\pm\phi(z_n=-c/2)$), 
\begin{eqnarray}\label{eq:twoplanes}
S=&\int \frac{d^2\xi}{(2\pi)^2} \frac14 \sum_{\pm}
\Big\{ 
\bm{\phi}_{\pm}
G_{\pm}
\bm{\phi}_{\pm} 
+ 2\bm{\phi}_{\pm} F_{\pm} 
\Big\} +\int d^2r S_F^{(2)}, \\
G_{\pm}=&\frac{\xi}{\sinh{c\xi}}\Big[
2\cosh{c\xi} - \alpha+\sqrt{\alpha^2-1} \mp1 \Big], \nonumber \\
F_{\pm}=&-\frac{8\pi|e|}{\sinh{c\xi}}  \co{ \sinh\frac{c\xi}2 \cosh{d\xi}  \\
-\cosh\frac{c\xi}2 \sinh{d\xi}  } \nonumber
\end{eqnarray}
This functional serves as a starting point for an analysis of the effects of non-linear screening in the two nearest planes assuming that all other metallic planes are in the linear screening regime. When the screening in the two nearest planes is linear,  $S_F=\frac14[ 2\kappa(\phi_+^2+\phi_-^2)
]$,  Eq.~(\ref{eq:twoplanes}) is solved with 
\begin{equation}\label{eq:linearscreening}
\phi^0_{\pm}=\frac{8\pi|e|}{\xi}
\co{ \sinh\frac{c\xi}2 \cosh{d\xi}  
-\cosh\frac{c\xi}2 \sinh{d\xi}  }
\times \frac{1}{\alpha+\sqrt{\alpha^2-1} \mp1} 
\end{equation}
which can also be obtained directly from Eq.~(\ref{eq:starting}). The planar screening length is controlled by the singularity of $\phi^0(\xi)$ in the complex plane of $\xi$ which is closest the real axis, i.e., one of the branch points $\xi^*$ which is a solution of  $\alpha(\xi^*)=\pm1$. The screening length approaches $c/\pi$  for  $\kappa c\gtrsim1$. 

To discuss the non-linear screening, we need to specify the compressibility part of the functional,  $S_F$, and variational functions. 
We take a simple model in which the non-linear compressibility in the metallic plane is a step function: when local screening charge density exceeds a certain threshold,  $\rho^*$, the metallic layer goes over into a local insulating (incompressible) state, $\delta \rho/\delta \phi=0$. In the cuprates, for instance, we have $\rho^* \sim 0.1$ holes per unit cell around optimal doping. In the absence of external charge, the local hole density is determined by chemical doping,  $\rho=0$. At high enough doping the copper-oxide plane is metallic and it is characterized by linear electrostatic response, $\delta\rho/\delta\phi = -\kappa/(2\pi |e|/\epsilon)$.
We note that when the threshold density $\rho^*$ is small a significant fraction of the electric field may leak to the next plane, $z_n=\pm3c/2$, and be strong enough for it to reach the threshold density as well, which will invalidate our assumption of all other planes being in the linear screening regime. The extension of  Eq.~(\ref{eq:twoplanes}) to allow non-linear screening in more than two nearest metallic planes is straightforward. Since the local charge density is defined via $S_F$,  this model requires $S_F = \kappa \phi^2 $ for $\phi<\phi^*$ and $S_F = \kappa\phi^*(2\phi-\phi^*)$ for $\phi>\phi^*$. Here $2\kappa\phi*=4\pi|e|\rho^*$. The effect of non-linear compressibility is that the screening charge is pushed out of the immediate vicinity of the external charge (when compared with the screening in the linear case), i.e., the screening cloud is larger. It is important that this does not affect the screening cloud at large distances where displaced charge is small and the screening is linear. This observation leads to the choice of our variational function: the electrostatic potential for a screening of a point charge in a non-linear metal is modelled with a screening potential in a linear metal due to a ``smeared'', finite size, external charge. Such variational function can be constructed starting with Eq.~(\ref{eq:linearscreening}). We further assume that the charge is only distributed in the direction parallel to the plane with charge density $\nu(r)$, i.e., the variational function is $\phi(\xi,z_n\!\!=\!\!\pm{c/2})\!=\!\nu(\xi)\phi_0(\xi,z_n\!\!=\!\!\pm{c/2})$ where $\phi_0$ is given by Eq.~(\ref{eq:linearscreening}). The simplest choice is a constant charge density over a disk of radius $R$, i.e., $\nu(r)=Q\theta(R-r)/(\pi{R}^2)$ where $\theta(x)$ is a step function. For the situation discussed below a non-uniform distribution $\nu(r)=Q\exp\{-r/R\}/(2\pi{R}^2)$ proves to be better. For the non-linear screening the density is not simply related (not proportional) to the local screening potential, thus $Q$ is an independent variational parameter. The interpolating property of our choice of variational  function is that in the linear screening case it is an exact screening potential with  $R=0$ and $Q=1$.
\begin{figure}[h]
\begin{centering}
\includegraphics{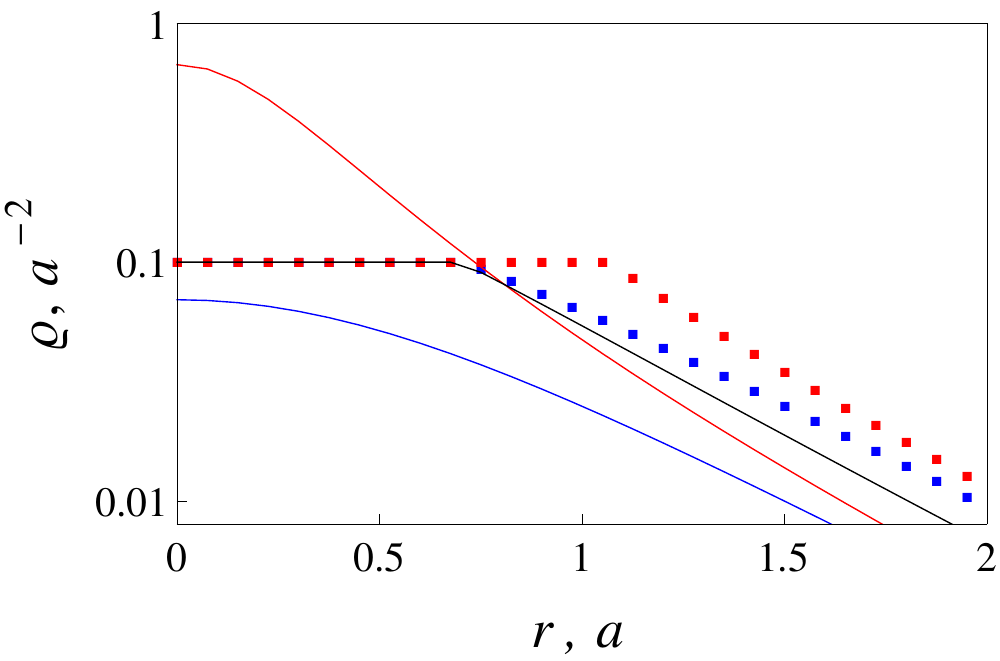}
\caption{Non-linear screening of the point charge $+1$ in a layered metal. We take $\kappa{c}=10$ and $c=1.5a$. The gray line is the non-linear screening density profile of an external charge in the center, $d=0$. The blue and red squares represent the screening density in the two planes for asymmetric position of an external charge, $d=c/4$. The solid red/blue lines show the linear screening density profile for the same geometry in the upper and lower metallic planes.}
\label{fig:Fig}
\end{centering}
\end{figure}
We now discuss screening of an external positive charge,  $+|e|$, in a layered metal which geometry which roughly corresponds to LSCO cuprates, presented in Fig. 1. The planar distance is in units of planar unit cell size, $a$, and the density is given in units of charge per planar unit cell. For a symmetric position of the external charge, the insulating region has a radius of  $0.68a$, and it contains about $-0.30$ charge (in units of  $+|e|$). Note that the for large values of linear compressibility ($\kappa{c}\gtrsim 1$) the density outside the insulating region decays rather slowly, at distances of about inter-plane distance, $c$, and the charge density decrease to a half of its value in insulating region at about $1a$ from the center. The total charge in each plane is $-0.5$,  ( $-0.48$ for linear screening in the same geometry). These numbers support our assumption that all other metallic planes are in the linear screening regime. The situation changes slightly if we shift the external charge away from the symmetric position, $d=c/4$, which more closely reflects the muon site in the LSCO lattice. The area of the insulating region in both planes increases in this asymmetric situation, $r_i=0.69a$ in one plane and $1.05a$ in the other. The total screening charge in each plane is, $-0.58$ and $-0.74$ respectively ($-0.26$ and $-0.69$ for linear screening in the same geometry). Compared to the symmetric external charge position, in this asymmetric situation the plane which is closer to the external charge is in the locally insulating state over a larger area and therefore, the effective, partially screened charge seen by the other, more distant plane, is also larger. The total charge of the two planes, $-1.33$, is greater than one, i.e., in this geometry the non-linear screening leads to an overscreening of the external charge in the two nearest metallic planes. The integrated screening charge in the infinite stack planes remains equal to the external charge. 

In conclusion, we have developed a minimal variational framework for the analysis of screening of a point charge in correlated layered metals such as cuprates in the metallic doping range. When applied to the screening of the muon in the LSCO cuprates, the calculation supports an earlier qualitative discussion in Refs.~[\cite{Shekhter08a, Dang10a}]. The method presented here may be used to study the charge distribution, rectification,  transistor action and other surface effects  between weakly doped Mott insulators and metals. 

\section{Acknowledgments}

We thank Vivek Aji and Albert Migliori for comments and suggestions. The work of GGGV and CMV was supported under the grant UC Lab fee research program 09-LR-01-118286-HELF.


\Bibliography{10}

\bibitem{KohnSham} W.~Kohn and L.~J.~Sham, \emph{Phys.~Rev.}~\textbf{140}, A1133 (1965).

\bibitem{Varma99a}
C. M. Varma, Phys. Rev. Lett. {\bf 83}, 3538 (1999).

\bibitem{Varma06a}
C. M. Varma, Phys. Rev. B {\bf 73}, 155113 (2006).

\bibitem{Sonier01a}
J. E. Sonier, J. H. Brewer, R. F. Kiefl, R. I. Miller,
G. D. Morris, C. E. Stronach, J. S. Gardner, S. R. Dunsiger,
D. A. Bonn, W. N. Hardy, R. Liang, and R. H. Heffner, Science {\bf 292}, 1692 (2001).

\bibitem{Sonier02a}
J. E. Sonier, J. H. Brewer, R. F. Kiefl, R. H. Heffner, K. F. Poon, S. L. Stubbs, G. D. Morris, R. I. Miller,
W. N. Hardy, R. Liang, D. A. Bonn, J. S. Gardner, C. E. Stronach, and N. J. Curro, Phys. Rev. B {\bf 66}, 134501 (2002).

\bibitem{MacDougall08a}
G. J. MacDougall, A. A. Aczel, J. P. Carlo, T. Ito, J. Rodriguez, P. L. Russo, Y. J. Uemura, S. Wakimoto, and G. M. Luke, Phys. Rev. Lett. {\bf 101}, 017001 (2008).


\bibitem{Sonier09a}
J. E. Sonier, V. Pacradouni, S. A. Sabok-Sayr, W. N. Hardy, D. A. Bonn, R. Liang, and H. A. Mook, Phys. Rev. Lett. {\bf 103}, 167002 (2009).  

\bibitem{Shekhter08a}
A. Shekhter, L. Shu, V. Aji, D. E. MacLaughlin, C. M. Varma,
Phys. Rev. Lett. {\bf 101}, 227004 (2008).

\bibitem{Dang10a}
H. T. Dang, E. Gull, A. J. Millis, 
Phys. Rev. B {\bf 81 }, 235124 (2010).


\bibitem{Ashcroft}
N. W. Ashcroft and N. D. Mermin, 
{\it Solid State Physics} (Thomson Learning, Inc., USA 1976).

\bibitem{Feynman}
R. P. Feynman, {\it Statistical Mechanics: A set of lectures},~(Westview Press, USA, 1998).

\endbib
\end{document}